\documentclass{article}
\usepackage{natbib}
\usepackage{icrctc07}

\title{Efficiency limits of diffusive shock acceleration}
\shorttitle{Efficiency limits of diffusive shock acceleration}
\authors{A. Meli$^{1}$ \& A. Mastichiadis$^{1}$}
\shortauthors{A. Meli et al}
\afiliations{$^1$Department of Physics, National University of Athens, 15783 Panepistimiopolis Zografos, Greece}
\email{ameli@phys.uoa.gr}

\abstract{It is well accepted today that diffusive acceleration in shocks results to the cosmic 
ray spectrum formation. This is in principle true for non-relativistic shocks, since there is a
detailed theory covering a large range of their properties and the resulting power-law 
spectrum, which is nevertheless not as efficient to reach the very high energies observed
in the cosmic ray spectrum. On the other hand, the cosmic ray maximum energy and 
the resulting spectra from relativistic shocks, are still under investigation and debate concerning their 
contribution to the features of the cosmic ray spectrum and the measured, or implied, cosmic ray 
radiation from candidate astrophysical sources. Here, we discuss the efficiency of the first order 
Fermi (diffusive) acceleration mechanism up to relativistic shock speeds, presenting Monte Carlo 
simulations.}

\begin{document}
\maketitle

\section{Introduction}

The notion that collisionless plasma shocks can accelerate charged particles into high energies
resulting into the observed cosmic ray spectrum, has been around for many years and by now it is widely 
accepted that this process can account for the majority of the energetic, non-thermal particle 
distributions that we infer in various astrophysical environments. The work of the late 70s by a number 
of authors (e.g. \cite{Krymsky77}, \cite{Bell78a}, \cite{Bell78b})  
established in principle the basic mechanism of particle diffusive acceleration in non-relativistic shocks. 
Since then, considerable work has been done analytically and numerically on the subject, however questions 
still need to be answered on this acceleration mechanism especially at relativistic shock speeds.


In the present paper we will present simulations from non-relativistic up to relativistic shock
speeds, in order to test the acceleration efficiency of the first order Fermi mechanism.
More precisely, these simulations include sub-luminal and super-luminal
shock acceleration with an aim to draw conclusions on the primary spectra of the relevant astrophysical 
sources such as Supernovae Remnants, Active Galactic Nuclei hot spots and Gamma Ray Bursts.

\section{Numerical simulations}

The purpose of using a Monte Carlo code is to simulate the particle transport equation, in our case 
for mild and relativistic shock velocities. The appropriate time independent Boltzmann equation is given
by

\begin{displaymath}
\Gamma(V+\upsilon\mu )\frac{\partial f}{\partial x}=\frac{\partial f}{\partial t}\arrowvert_{c} 
\end{displaymath}

where $V$ is the fluid velocity, $\upsilon$ the velocity of the particle, $\Gamma$ the Lorentz factor
of the fluid frame, $\mu$ the cosine of the particle's pitch angle and $\frac{\partial f}{\partial t}|_{c}$
the collision operator. The frames of reference used in the code
are the local fluid frame, the normal shock frame  and
the de Hoffman-Teller frame. Pitch angle is measured in the local fluid frame,  while the value $x$ is
the distance of the particles to the shock front, where the shock is assumed to be placed at $x=0$.
Specifically, in order for the above equation to be solved by the Monte Carlo technique, we assume that i) the collisions represent scattering in pitch
angle and ii) the scattering is elastic in the fluid frame.
\begin{figure}[t]
\begin{center}
\includegraphics [width=6.8cm]{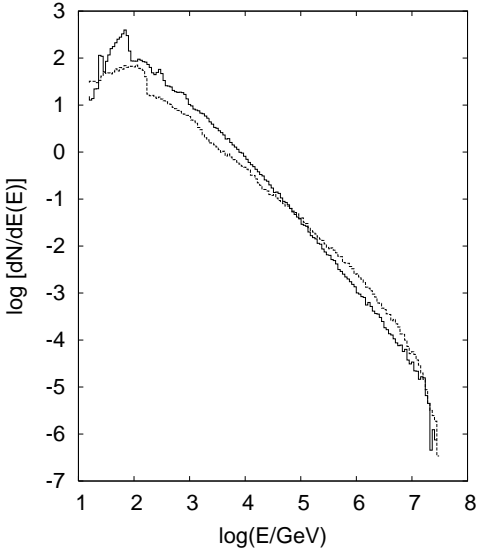}
\end{center}
\caption{Two sub-luminal shock spectra, one for a non-relativistic shock of a
gamma factor equal to 1.0001 (i.e. $V$=0.01c) and the other for a gamma equal to
5.0 (i.e. $V$=0.98c) with an inclination angle of 89 and 39 degrees respectively. One can see
the almost perfect superposition of the spectra, showing the equivalence of acceleration efficiency between the 
non-relativistic \textit{perpendicular} shock and the relativistic sub-luminal oblique one. }
\end{figure}
\begin{figure}[h]
\begin{center}
\includegraphics [width=7cm]{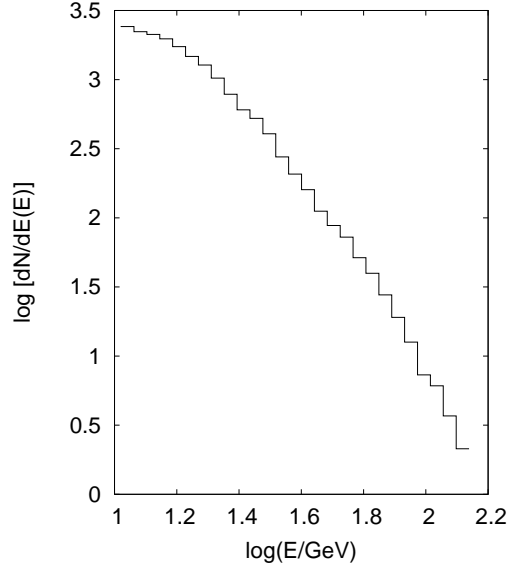}
\caption{Super-luminal shock spectrum for a gamma factor of 10 at 89 degrees. }
\end{center}
\end{figure}
We begin the simulation by injecting $10^{5}$ particles far upstream.
First order Fermi (diffusive) acceleration can be simulated provided that the particles' guidance center 
undergoes multiple scatterings with the assumed magnetized media and in each shock crossing the particles 
gain an amount of energy. 
The mean free path $\lambda$ is calculated in the respective fluid frames (i.e. upstream or downstream)
assuming a momentum dependence to the mean free path of the particle such as, $\lambda=\lambda_o \cdot p$, where $\lambda_o=10 \cdot r_g$, which is related to the spatial diffusion coefficient, $\kappa$. 
Specifically for oblique shocks, in the shock normal 
which lies in the $x$ direction, the diffusion coefficient $\kappa$,  equals to $\kappa_{\|}cos^{2} \psi$, 
where $\kappa_{\|}=\lambda v/3$ and $\psi$ is the inclination of the shock to the magnetic field lines. 
A pitch angle diffusion algorithm is used (e.g. see \cite{MeliQb}).
All results are given at the downstream side of the shock frame.
In general, flow into and out of the shock discontinuity
is not along the shock normal, but a transformation is possible into the shock frame to render
the flows along the normal (e.g. \cite{BegelKirk90}) and for simplicity we assume that such
a transformation has already been made.
During our simulations continuous Lorentz transformations are performed from the local fluid rest
frames to the shock frame to check for shock crossings.\\
Furthermore, for the super-luminal shock conditions (see \cite{MeliQb}) we consider a helical trajectory 
motion of each test-particle of momentum $p$,  in the fluid frames upstream or downstream 
respectively, where the velocity coordinates ($v_x, v_y, v_z$) of the particle are calculated in a three dimensional space. In principle, we follow the helix trajectory of each particle in time $t$,
where $t$ is the time from detecting the shock intersection at $x$, $y$, $z$. 
Further details of this method can be found in the work mentioned above.
All particles leave the system if they escape far downstream
at the spatial boundary $r_b$, or alternatively, if they reach a specified maximum energy $E_{max}$, for
computational convenience, even though other physical parameters describing particle escape or energy loss
would need to be taken into account in more realistic situations.
The downstream spatial boundary required, can be initially estimated from the solution of the convection-diffusion
equation in a non-relativistic, large-angle scattering approximation in the downstream plasma which gives the
chance of return to the shock as, $exp(-V_{2}r_{b}/x_i)$.
In fact, we have performed many runs with different spatial boundaries 
to investigate the effect of the size of the acceleration region on the spectrum, as well as to find a region where the spectrum is size independent. We note that in the pitch angle diffusion case, the inherent 
anisotropy due to the high downstream sweeping effect may greatly modify this analytical estimate.

\begin{figure}[t]
\begin{center}
\includegraphics [width=7cm]{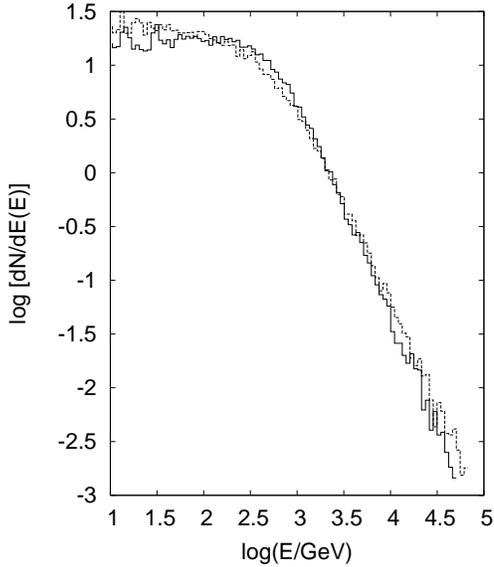}
\caption{Two super-luminal shock spectra for a gamma factor of 500, one for an inclination angle of 50 
degrees and the other for an inclination of 89 degrees. The spectra reach a maximum of $10^{14}$eV for protons. This cut-off in the energy is due to the stream flow, sweeping the particles downstream limiting their 
chance of returning back to the shock. }
\end{center}
\end{figure}

\begin{figure}[h]
\begin{center}
\includegraphics [width=7.0cm] {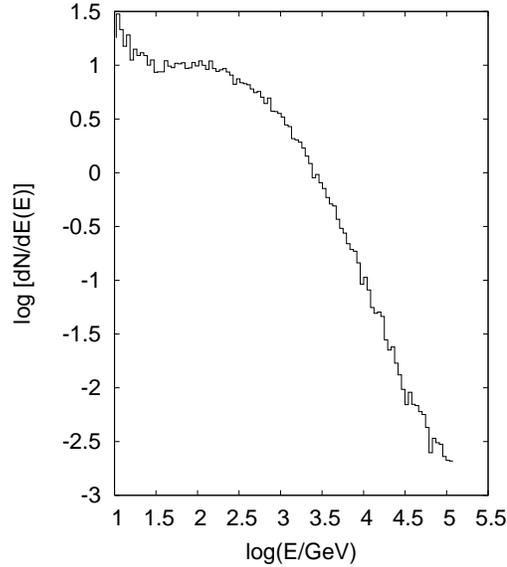}
\caption{A super-luminal shock spectrum for a gamma factor of 900 at 75 degrees. One sees (simulated
spectra for other shock inclinations are comparable) the smooth power-law formation of the spectrum. 
Nevertheless, even for such high Lorentz factor the super-luminal shocks do not seem  to be as efficient 
as the sub-luminal ones (e.g. \cite{MeliQb}), since the simulations showed that almost all of the particles
are advected downstream just after a shock cycle.}
\end{center}
\end{figure}

\section{Results}

In this section we present results of the simulations shown in the figures 1-4.\\
Two sub-luminal shock spectra are shown in figure 1, one for a non-relativistic shock of a
gamma factor equal to 1.0001 (i.e.~$V$=0.01c) and the other for a gamma equal to 
5.0 (i.e.~$V$=0.98c) with an inclination angle of 89 and 39 degrees respectively. One can see 
the almost perfect superposition of the spectra, showing the equivalence of the acceleration efficiency 
between the non-relativistic \textit{perpendicular} shock and the relativistic sub-luminal oblique one. The
above correlation of efficiency between non-relativistic perpendicular shocks and mild relativistic oblique ones, 
applies to many other shock spectra we simulated, as long as the non-relativistic shocks are nearly 
perpendicular.
For the case shown here, the maximum energy corresponds to $\sim 10^{7.5}$GeV for protons. 
Adding to the above, \cite{MeliBierm06} have shown that non-relativistic perpendicular shocks,
are efficient accelerators for cosmic rays reaching  energies
as high as $\sim 10^{17}$eV, since the acceleration in those shocks seems to be faster 
under certain diffusion conditions \cite{Jok87}. Also as \cite{MeliQb} have shown, sub-luminal highly 
relativistic shocks are the most efficient accelerators comparing to the above, resulting in an average cosmic 
ray energy gain of $\sim \Gamma^{3.4}$ after two complete shock cycles. For this case, $\Gamma$ is the 
boost factor of the shock, while in the simulations the test-particles are considered already relativistic 
with an initial Lorentz gamma, $\gamma=\Gamma$+100. \\
One further result of the simulations, is that for oblique shocks below a Lorentz factor of
$\Gamma$=5 (i.e. $V$=0.98c) as higher the inclination of the shock as greater its efficiency to accelerate 
particles up to $10^{17}$eV (i.e. protons, iron nuclei). \\
In figure 2  we show a super-luminal shock spectrum for a gamma factor of 10 at 89 degrees.
In figure 3 one sees two super-luminal shock spectra for a gamma factor of 500, one for an inclination 
angle of 50 degrees and the other for an inclination of 89 degrees. Simulations show that the inclination 
of the shock in the super-luminal cases does not change the results. The cut-off is due to the flow, 
sweeping the particles downstream limiting their chance to return back to the shock, diminishing the 
chances for these shocks to be efficient cosmic ray accelerators. In figure 4, a super-luminal shock 
spectrum for shock gamma factor of 900 at 75 degrees is shown. One sees the smooth 
power-law formation of the above spectra (in contrast to the sub-luminal spectra formations in 
\cite{MeliQb}). Nevertheless, even for as high Lorentz gamma as 900, the accelerated particles do not 
reach very high energies (e.g. $10^{14}$eV for protons) since almost all of them are advected downstream, 
after their helix trajectory performs a shock-cycle crossing. All the simulated super-luminal shock spectra
can be well fitted by a power-law with a spectral index ranging between $\sim 2.0-2.3$ in contrast to the 
work of \cite{MeliQb} concerning sub-luminal relativistic shocks. The latter spectra 
appear flatter, at the highest speeds with a characteristic plateau-like structure, since first, a highly relativistic shock catches the particles up within less a shock cycle, which does not allow sufficient time for isotropisation and second, because relativistic shocks are sensitive to the applied particle scatter model \cite{MeliBecker07} .

\section{Conclusions}

We discussed the mechanism of diffusive shock acceleration, presenting 
simulations in order to test the efficiency of sub-luminal and super-luminal relativistic shocks 
in a quantitative comparison to non-relativistic ones.  
It is certain that super-luninal shocks are not as efficient first order Fermi cosmic ray accelerators 
as the sub-luminal ones. Nevertheless, the super-luminal shocks can be well fitted by a power-law with a 
spectral index ranging between $\sim 2.0-2.3$ in contrast to the relativistic sub-luminal ones which give 
flatter spectra. Specifically, the super-luminal shocks are efficient in accelerating particles up to a 
maximum energy of around $10^{14}$eV (i.e. protons or iron nuclei). On the other hand, the sub-luminal 
relativistic oblique shocks are better accelerators concerning a maximum energy above $ 10^{18}$eV, as shown 
in similar works as well, but the spectral features are not as smooth as in super-luminal ones (which are given by a 'clean' power-law), indicating a strong connection to the kinematical details of the particle diffusion while crossing the relativistic shock front.
Interestingly, it is shown that perpendicular non-relativistic shocks seem to be as efficient accelerators as 
sub-luminal mild relativistic ones. There is work under way concerning the initial parameters
which could affect the shock behaviour in highly relativistic speeds.

\section{Acknowledgements}
The project is funded by the European Social Fund and National Resources (EPEAEK II) PYTHAGORAS.

\bibliography{libros}
\bibliographystyle{plain}

\end{document}